\documentclass[journal]{IEEEtran}
\usepackage{cite}
\usepackage{amsmath}
\usepackage{graphicx}
\ifCLASSINFOpdf

\else

\fi

\begin{document}
	\title{Data-driven decomposition of brain dynamics with principal component analysis in different types of head impacts}
	\author{Xianghao~Zhan,
		Yuzhe~Liu,
		Nicholas~J.~Cecchi,
		Olivier~Gevaert,
		Michael~M.~Zeineh,
		Gerald~A.~Grant,
		and~David~B.~Camarillo,~\IEEEmembership{Member,~IEEE}
		\thanks{X. Zhan, Y. Liu, N. Cecchi, and D. Camarillo are with the Department of Bioengineering, Stanford University, Stanford, 94305, USA. (Corresponding Author: Yuzhe Liu e-mail: yuzheliu@stanford.edu)}
		\thanks{O. Gevaert is with the Department of Biomedical Data Science and Stanford Center for Biomedical Informatics Research, Stanford University, Stanford, 94305, USA}
		\thanks{M. Zeineh is with the Department of Radiology, Stanford University, Stanford, 94305, USA}
		\thanks{G. Grant is with the Department of Neurosurgery, Stanford University, Stanford, 94305, USA}
		\thanks{Manuscript received ; revised .}}
	
	\maketitle

	\begin{abstract}
		Objective: Strain and strain rate are effective traumatic brain injury predictors. Kinematics-based models estimating these metrics suffer from significant different distributions of both kinematics and the injury metrics across head impact types. To address this, previous studies focus on the kinematics but not the injury metrics. 
		We have previously shown the kinematic features vary largely across head impact types, resulting in different patterns of brain deformation. This study analyzes the spatial distribution of brain deformation and applies principal component analysis (PCA) to extract the representative patterns of injury metrics (maximum principal strain (MPS), MPS rate (MPSR) and MPS$\times$MPSR) in four impact types (simulation, football, mixed martial arts and car crashes).
		Methods: We apply PCA to decompose the patterns of the injury metrics for all impacts in each impact type, and investigate the distributions among brain regions using the first principal component (PC1). Furthermore, we developed a deep learning head model (DLHM) to predict PC1 and then inverse-transform to predict for all brain elements.
		Results: PC1 explained $>80\%$ variance on the datasets. Based on PC1 coefficients, the corpus callosum and midbrain exhibit high variance on all datasets. We found MPS$\times$MPSR the most sensitive metric on which the top 5\% of severe impacts further deviates from the mean and there is a higher variance among the severe impacts. Finally, the DLHM reached mean absolute errors of $<0.018$ for MPS, $<3.7 s^{-1}$ for MPSR and $<1.1 s^{-1}$ for MPS$\times$MPSR, much smaller than the injury thresholds.
		Conclusion: The brain injury metric in a dataset can be decomposed into mean components and PC1 with high explained variance.
		Significance: The brain dynamics decomposition enables better interpretation of the patterns in brain injury metrics and the sensitivity of brain injury metrics across impact types. The decomposition also reduces the dimensionality of DLHM.
	\end{abstract}
	
	
	\IEEEpeerreviewmaketitle
	
	\section{Introduction}
	Traumatic brain injury (TBI) is a major public health concern across the globe. It has been approximated that there are more than 55 million prevalent TBI cases, and over 20 million new cases of TBI are estimated to occur in developed countries annually \cite{james2019global}. Each year, more than 1 million people suffer TBI in the United States alone \cite{dompier2015incidence}. The onsets of TBI can be attributed to head impacts from multiple sources, including but not limited to accidental falls, traffic accidents, blasts, various contact sports, and domestic abuse \cite{caswell2017characterizing,cecchi2019head,corrigan2003early,hernandez2015six,o2020dynamic,versace1971review,wilcox2014head}. While single-incidence TBI can immediately cause loss of consciousness and disabilities, repetitive incidence of milder TBI can also lead to long-term cognitive deficits, even neurodegenerative diseases and chronic traumatic encephalopathy \cite{doherty2016blood,dekosky2013acute}. 
	
	In recent decades, with the development of biomechanics modeling technology, many effective predictors of TBI based on brain dynamics (such as strain, strain rate, and their combination have been proposed to detect TBI for better protection and prevention. Physiologically, after the head gets impacted, the acceleration and deceleration of the head result in brain deformation. The high magnitude and rate of brain deformation can result in multiple pathologies, such as traumatic axonal injury (TAI) and blood-brain-barrier disruption \cite{bain2000tissue, bar2016strain, donat2021biomechanics, cater2006temporal,fahlstedt2015correlation,hajiaghamemar2020head,hajiaghamemar2020embedded}. Therefore, these injury metrics can be used as the predictors of TBI. The state-of-the-art method to calculate these injury metrics from kinematics is finite element modeling (FEM) \cite{kleiven2007predictors,mao2013development}, in which the brain is modeled by more than thousands elements. However, as the whole-brain dynamics are determined by the head kinematics and brain physics, the underlying degrees of freedom (DOF) may be much lower than the number of FEM brain elements. Therefore, it is hypothesized that there is spatial co-variation in the injury metrics among different brain elements, and it is possible to find the reduced-order interpretation for whole-brain strain and strain rate.
	
	Previously, researchers have put in efforts to find the reduced-order decomposition of kinematics and brain deformation in the biomechanics research of TBI. For example, a data-driven emulator was developed to simulate the kinematics of head impacts with the principal components found by principal component analysis (PCA) (e.g., 15 principal components for angular velocity simulation) \cite{arrue2020low}. Additionally, the dynamic mode decomposition was adopted to extract the deformation modes \cite{laksari2018mechanistic}, and a convolutional-neural-network-based human head model \cite{ghazi2021instantaneous} was combined with a pre-computed brain \cite{ji2015pre} response dataset to find the effective kinematics that yields similar brain deformation for the actual kinematics \cite{ghazi2021effective}. The previous studies focus on the temporal co-variation in the kinematics \cite{arrue2020low}, the temporal co-variation of brain deformation \cite{laksari2018mechanistic}, and the co-variation in the relationship between kinematics and deformation \cite{ghazi2021effective}. There has been no study insofar that focuses on the spatial co-variation of the peak values in a group of head impacts and particularly the spatial co-variation across different brain elements for a specific dataset.
	
	In this study, we used principal component analysis to investigate the decomposition of three injury metrics for TBI based on brain-dynamics: maximum principal strain (MPS), maximum principal strain rate (MPSR), and the product of MPS and MPSR (MPS$\times$MPSR). With a total of 3,161 impacts from head model simulations, college football, mixed martial arts and car crashes, we have shown that the injury metrics can be decomposed into the mean component (PC0) and the first principal component (PC1), which can generally explain more than 80\% variance for each impact dataset. With the decomposition in hand, by predicting the values on PC1 with the PCA inverse-transformation, we were able to develop a deep learning head model (DLHM) to predict the whole-brain MPS, MPSR and MPS$\times$MPSR accurately. Additionally, by analyzing the data variance on PC1, we demonstrated that the metric MPS$\times$MPSR, and brain regions such as the corpus callosum, bear higher sensitivity for severe impacts.
	
	\section{Methods}
	\subsection{Datasets: head kinematics and brain finite element model}
	In this study, to cover a broader range of head impacts, we included 3,161 impact data from four different impact sources: 1) 2,130 football-like impacts simulated by a finite element model of a Hybrid III anthropomorphic test device headform (ATD) (labeled as HM for head model) \cite{giudice2019development,zhan2021rapid}; 2) 302 college football impacts measured by the Stanford Instrumented Mouthguard (labeled as CF) \cite{camarillo2013instrumented,liu2020validation,liu2021time}; 3) 457 mixed martial arts impacts measured by the Stanford Instrumented Mouthguard (labeled as MMA) \cite{tiernan2020concussion,o2020dynamic}; 4) 272 reconstructed impacts from the National Association for Stock Car Auto Racing (labeled as NASCAR) \cite{zhan2021classification,zhan2021predictive}.
	
\begin{figure}[htbp]
		\centering
		\includegraphics[width=0.99\linewidth]{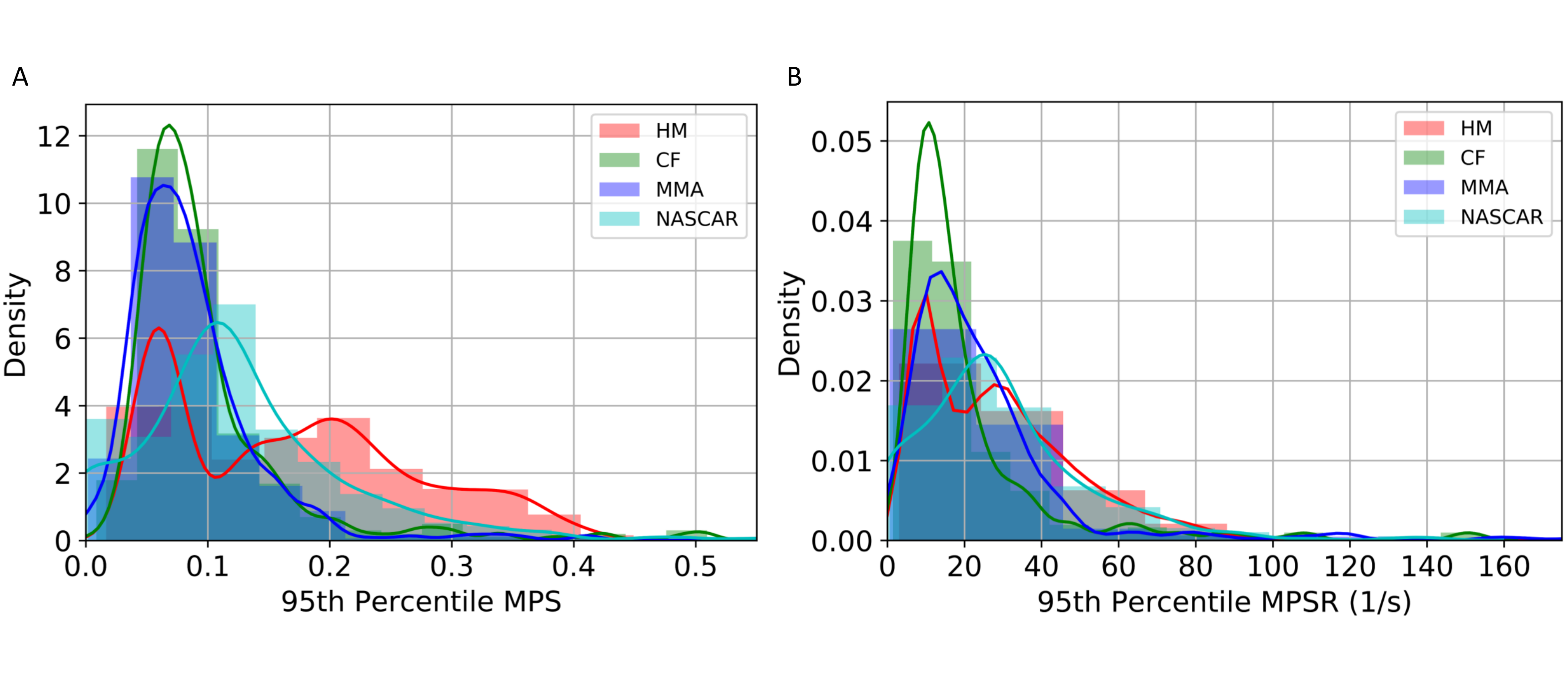}
		\caption{The distribution of the MPS, and MPSR across various datasets. The distribution of the 95th percentile MPS (A) and the 95th percentile MPSR (B).}
		\label{dist}
		\vspace{-5mm}
	\end{figure}

	To calculate the brain dynamics in this study, we used the KTH FE model (Stockholm, Sweden) \cite{kleiven2007predictors}, which is a validated FE head model \cite{kleiven2006evaluation,zhou2019reanalysis}. With 4,124 brain elements, the KTH FE model models the brain, skull, scalp, meninges, falx, tentorium, subarachnoid cerebrospinal fluid (CSF), ventricles, and 11 pairs of the largest bridging veins. The model has been validated by the experimental data of brain-skull relative motion \cite{kleiven2006evaluation}, intracranial pressure \cite{kleiven2006evaluation}, and brain strain \cite{zhou2019reanalysis}. In this study, the MPS and MPSR for each of the 4,124 brain elements were calculated by KTH models. As a result, the entire datasets can be represented by three 3,161 $\times$ 4,124 matrices for MPS, MPSR and MPS$\times$MPSR respectively. The distribution of the MPS and MPSR modeled by KTH FE model for these datasets is shown in Fig. \ref{dist}.
	
	\subsection{Brain dynamics decomposition with principal component analysis}
	As the whole-brain dynamics were modeled based on the kinematics input, it is possible that the whole-brain dynamics in a specific dataset can be modeled with DOF smaller than the number of brain elements (i.e., 4,124), considering the spatial co-variations among different brain elements. To find the potential reduced-order representation of the whole-brain dynamics for the specific dataset, we leveraged the principal component analysis (PCA), which is an unsupervised dimensionality reduction method that takes co-variation into consideration. PCA seeks the reduced-order representation of data by finding the directions where the variance is maximized. In this study, each impact of the four datasets can be denoted as ${\vec{x^{(i)}}; i = 1, 2, ..., n}, \vec{x^{(i)}} \in \rm I\!R^{4124}$ where $n$ is the number of impacts for a specific dataset (e.g., $n=302$ for dataset CF, $n=457$ for dataset MMA). To begin with, the data were centered by removing the mean vector: $\Tilde{x}^{(i)} = \vec{x^{(i)}} - \vec{\mu}$, where $\vec{\mu} \in \rm I\!R^{4124}$ is the mean vector across different samples in a specific dataset. The mean vector represents the center of the distribution of a particular impact dataset and injury metric, which was regarded as principal component zero (PC0) in this study. In order to find the set of basis vectors $\vec{v_1}, \vec{v_2}, ..., \vec{v_k} \in \rm I\!R^{4124}, k<4124$, where the data resolved onto the directions represented by the basis vectors are with the highest variance, singular value decomposition (SVD) is performed on the data matrix $\Tilde{X}$: 
	\begin{equation}
	\Tilde{X} = U \times \Sigma \times V^T
	\end{equation}
	
	Upon ranking the singular values from large to small, the associated vectors $\vec{v_1}, \vec{v_2}, ..., \vec{v_k} \in \rm I\!R^{4124}, k<4124$ in $V$ represent the projection directions with the widest data spread (i.e., highest variance). These vectors are also referred to as the principal components (PCs) or the PC coefficients. The transformed low-dimensional data along each direction: $\vec{\alpha_1}, \vec{\alpha_2},..., \vec{\alpha_k} \in \rm I\!R^{n}$ can be computed by solving the following optimization problem:
	\begin{equation}
	\mathrm{argmin}_{\vec{\alpha_1}, \vec{\alpha_2},..., \vec{\alpha_k}} ||\Tilde{X}-\Sigma_j^k \vec{\alpha_j} \times \vec{v_j}^T||^2
	\end{equation}
	This results in the projection of the high-dimensional data onto these basis vectors: $\vec{\alpha_j} = <\vec{v_j}, \Tilde{X}>$. With PCA, we can reduce the dimensionality of the data from 4124-D to $k$-D. To determine how many orders suffice, we firstly calculated the explained variance ($R^2$) with varying values of $k$. Upon determining $k$, the brain dynamics matrices can be changed from $n \times 4124$ matrices to $n \times k$ matrices. With the PCA, the reduced-order approximation of the original data matrix $X$ can be represented as:
	\begin{equation}
	X = \vec{1} \times \vec{\mu}^T + \Sigma_j^k \vec{\alpha_j} \times \vec{v_j}^T
	\end{equation}
	where the $\vec{1} \in \rm I\!R^{n}$ denotes a vector with ones as its elements. The first term in this approximation equation denotes the PC0 contributions to the injury metrics.
	
	\subsection{Interpretation of PC1 in severe head impacts}
	With the reduced-order principal components to represent the injury metrics in a head impact dataset, we were able to summarize the injury metrics (MPS/MPSR/MPS$\times$MPSR) using components in PCA (PC0, PC1, etc). For each dataset, we found that PC1 can explain more than 80\% variations among impacts, which means that the distribution of metrics can be represented by PC0 and PC1 (See Results Section). Since PC0 denotes the mean components, which is the same across different impacts. Therefore, PC1, representing the difference among impacts, contributes more to the injury metrics in severe head impacts. To confirm this, we firstly ranked the severity of the impact by calculating the 95th percentile MPS/MPSR/MPS$\times$MPSR for each impact and then ranked the impacts and selected the top 5\% most severe impacts for this analysis. The reason why we focused on the top 5\% severe impacts was that we emphasize the modeling of impacts that are more dangerous while the majority of the simulated impacts and on-field impacts were mild impacts. Then, the contribution of different principal components was quantified according to the following formula:
	\begin{equation}
	\mathrm{c_i} = \Sigma_{j}^{4124}\frac{P(i,j)}{T(i,j)}, i = 1,2,...,n
	\end{equation}
	where $P(i,j)$ denotes the brain dynamics metric value of the $i$-th impact and $j$-th brain element that is contributed by one principal component (PC0, PC1, etc.) and $T(i,j)$ denotes the reference brain dynamics metric value of the $i$-th impact and $j$-th brain element given by the KTH FE model. For example, the $P(i,j)$ from PC0 is $\mu \in \rm I\!R^{4124}$ and the $P(i,j)$ from PC$k$ is $\vec{\alpha_{ki}} \times \vec{v_{kj}}$. In this manner, the MPS, MPSR and MPS$\times$MPSR for each brain element can be decomposed into the contributions from different components and the contributions can be quantified by the ratios.
	
	In addition to quantifying the overall contribution of the principal components, we also quantified the contribution to the injury metrics in different brain regions that have been modeled by the KTH FE model: brainstem (BS), corpus callosum (CC), cerebellum (CL), gray matter (GM), midbrain (MB), thalamus (TH) and white matter (WM).
	
	\subsection{Sensitivity of different injury metrics}
	Brain injury metrics are used to decide the severity of the impacts. The sensitivity of the injury metrics is important in differentiating more severe impacts and less severe impacts and the higher variance in the injury metrics can benefit the prediction of injury or non-injury. With PCA and the contribution calculated in the previous section, we analyzed the sensitivity of the different brain injury metrics (MPS, MPSR and MPS$\times$MPSR) by the values and the variances of the PC1 contribution for the top 5\% most severe impacts across different datasets. The reason we quantified the sensitivity with PC1 contribution was: firstly, PC0 and PC1 generally explained more than 80\% variance in our datasets; secondly, PC0 contribution (mean) is a constant shared by all impacts in a dataset while PC1 contribution represents the majority of variance in the brain injury metrics. As we decomposed the brain injury metrics into PC0 contribution and PC1 contribution, the larger the PC1 contribution, the more the most severe impacts deviate from the mean, which indicates the higher sensitivity of a specific brain injury metric. The larger variance on PC1 contribution can also indicate that the injury metrics can better separate the top 5\% most severe impacts.
	
	It is worth noting that instead of directly quantifying the injury metric sensitivity with the values and the variance on PC1 (i.e., $\vec{\alpha_1}$, $var(\vec{\alpha_1})$), to account for the different units of the three injury metrics, we used the PC1 contribution calculated in the previous section, which can be viewed as the normalized PC1 contribution by the ground truth MPS/MPSR/MPS$\times$MPSR.

	\subsection{Application in deep learning head models}
	The state-of-the-art FE simulation can not provide real-time monitor of the injury metrics, while DLHM can act as function approximators of FE models to compute injury metrics such as MPS, MPSR and MPS$\times$MPSR \cite{zhan2021rapid}. However, the DLHMs, with thousands of output units for the brain elements, require large quantities of training data. The decomposition of brain dynamics with PCA can reduce the dimensionality of output and therefore simplify the training objective in the development of DLHMs. To verify this, we leveraged the PCA models to get the low-dimensional brain dynamics representations, developed DLHMs to predict the value on the $k$ PCs based on 510 kinematics features \cite{zhan2021rapidly} (with $k$ determined in the first results section), and reconstructed the whole-brain MPS, MPSR and MPS$\times$MPSR with the accuracy evaluated by several metrics. The details are shown as follows:
	
	The 510 kinematics features were extracted from the four types of kinematics describing the head movement: linear acceleration at the head center of gravity $a(t)$, angular velocity $\omega(t)$, angular acceleration $\alpha(t)$ and angular jerk $j(t)$. Among them, $a(t)$ and $\omega(t)$ were measured by the sensors while $\alpha(t)$ and $j(t)$ were calculated from $\omega(t)$ with a five-point stencil numerical derivative equation. The features include both the temporal features such as the maximum values and integral values, and the spectral densities within 19 frequency windows\cite{zhan2021clustering}, because these features were found to show high predictability of brain strain and strain rate in our previous work \cite{zhan2021clustering,zhan2021rapid}.
	
	The DLHM was developed on two datasets: 1) dataset HM: the 2,130 head model simulated impacts, and 2) dataset MIX: the mixture of all four datasets with 3,161 impacts. To develop the models, we firstly partitioned the datasets into a training set (70\%), a validation set (15\%) and a test set (15\%), according to the same protocol mentioned in our previous studies \cite{zhan2021rapid,zhan2021clustering}. The training set was used to train the model parameters, the validation set was used to tune the hyperparameters (such as the number of neurons in each layer, the number of epochs to train the models), and the test set was used to evaluate the model accuracy. In order to get robust results, we randomly partitioned the datasets 20 times and did 20 parallel experiments.
	
	Then, we trained the PCA dimensionality reduction model on the training set and recorded PC0 and the principal components (PC1, PC2, ..., PC$k$). A five-layer deep neural network (besides the input and output layers) was developed to learn the mapping from the kinematics features to the values on the $k$ principal components. The second and the fourth layers were the dropout layers with a dropout rate of 0.5. L2 regularization was used together with the dropout to further regularize the DLHMs and avoid overfitting. During the training, to further optimize the accuracy, we adopted the same data augmentation strategy used in our previous studies: adding slight Gaussian noises to generate shadow samples \cite{zhan2019feature,liu2021boost}.
	
	Upon training the model, in the prediction stage, we reconstruct the 4124-D MPS, MPSR and MPS$\times$MPSR by taking the inverse-transform on the predicted values on the $k$ PCs. To evaluate the model accuracy, we calculated the mean absolute errors (MAE) and coefficient of determination ($R^2$) between the predicted MPS/MPSR/MPS$\times$MPSR and the reference values given by the KTH FE model.
	
	\section{Results}
	\subsection{Determination of the orders for brain dynamics}
\begin{figure}[htbp]
		\centering
		\includegraphics[width=0.99\linewidth]{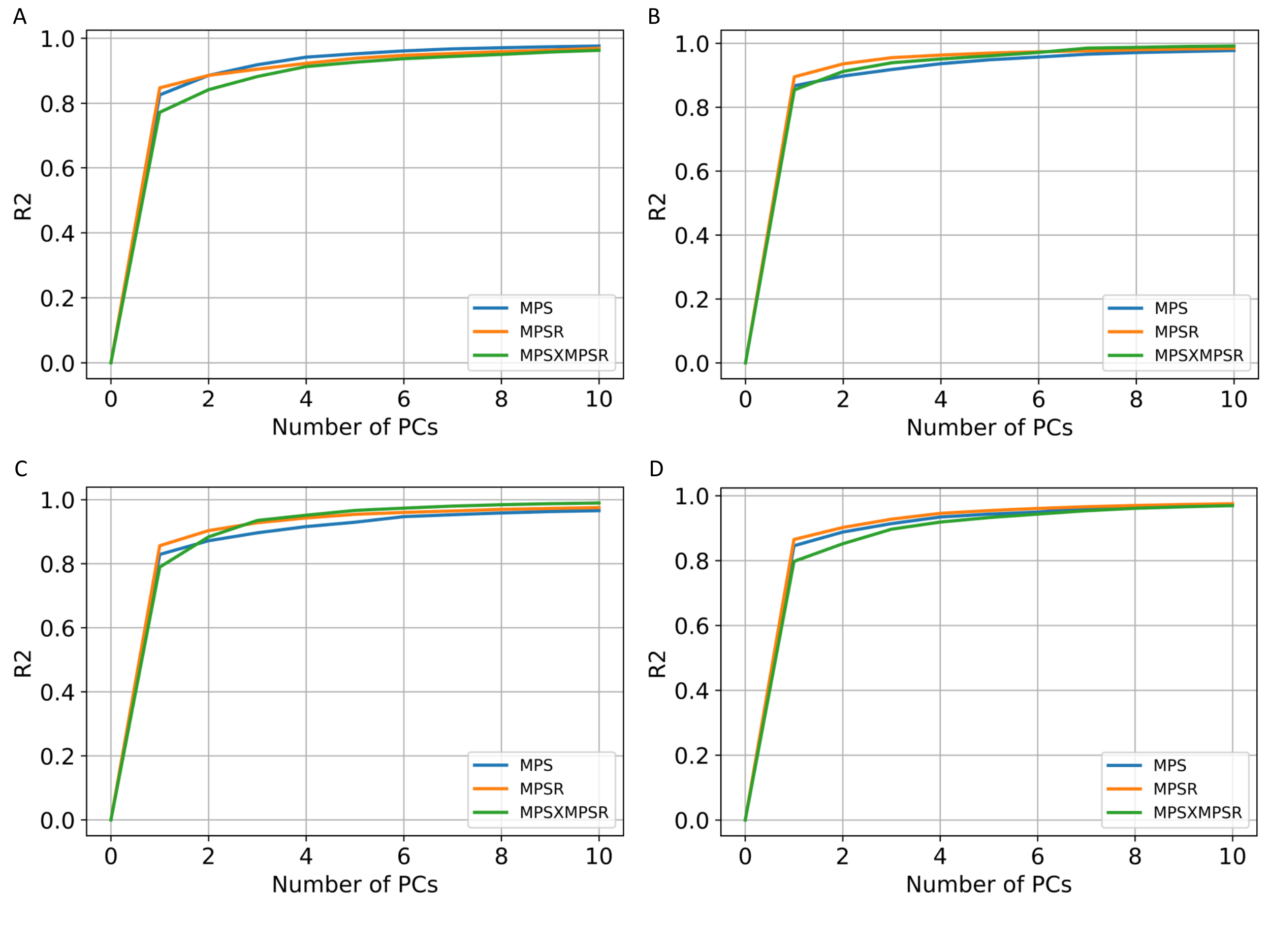}
		\caption{The cumulative explained variance with different orders of principal components. The explained variance for the three injury metrics on dataset HM (A), dataset CF (B), dataset MMA (C) and dataset NASCAR (D).}
		\label{explained variance}
		\vspace{-5mm}
	\end{figure}

	In the determination of the orders for the three injury metrics, we first computed the explained variance based on the PCA. According to the results shown in Fig. \ref{explained variance}. The PC0 and PC1 are generally able to explain more than 80\% variance in each dataset. We have also tested the explained variance when the four datasets are combined and the results also showed that more than 75\% variance can be explained by PC0 and PC1. Therefore, we chose PC0 and PC1 as the reduced-order representations of the three injury metrics. To visualize the contributions, we plotted the reference values of the injury metrics, the contribution of PC0, and the contribution of both PC0 and PC1 in Fig. \ref{example visualization} on dataset HM as an example (results on other datasets shown in Supplementary Fig. 1-3). The results indicate that the PC0 and the PC1 sufficiently explain the majority of the variation of the entire dataset. Visually, although the PC0 cannot explain the details in the reference value heatmaps, the addition of PC1 enables the explanation of much of the details in the heatmaps. The Bland Altman plots for the reconstruction error distribution on four datasets and three injury metrics are shown in Supplementary Fig. \ref{reconstruction error}, which indicates that the PC0 and PC1 can lead to the accurate injury metrics values for the majority of data.
	
	\begin{figure*}[htbp]
		\centering
		\includegraphics[width=0.99\linewidth]{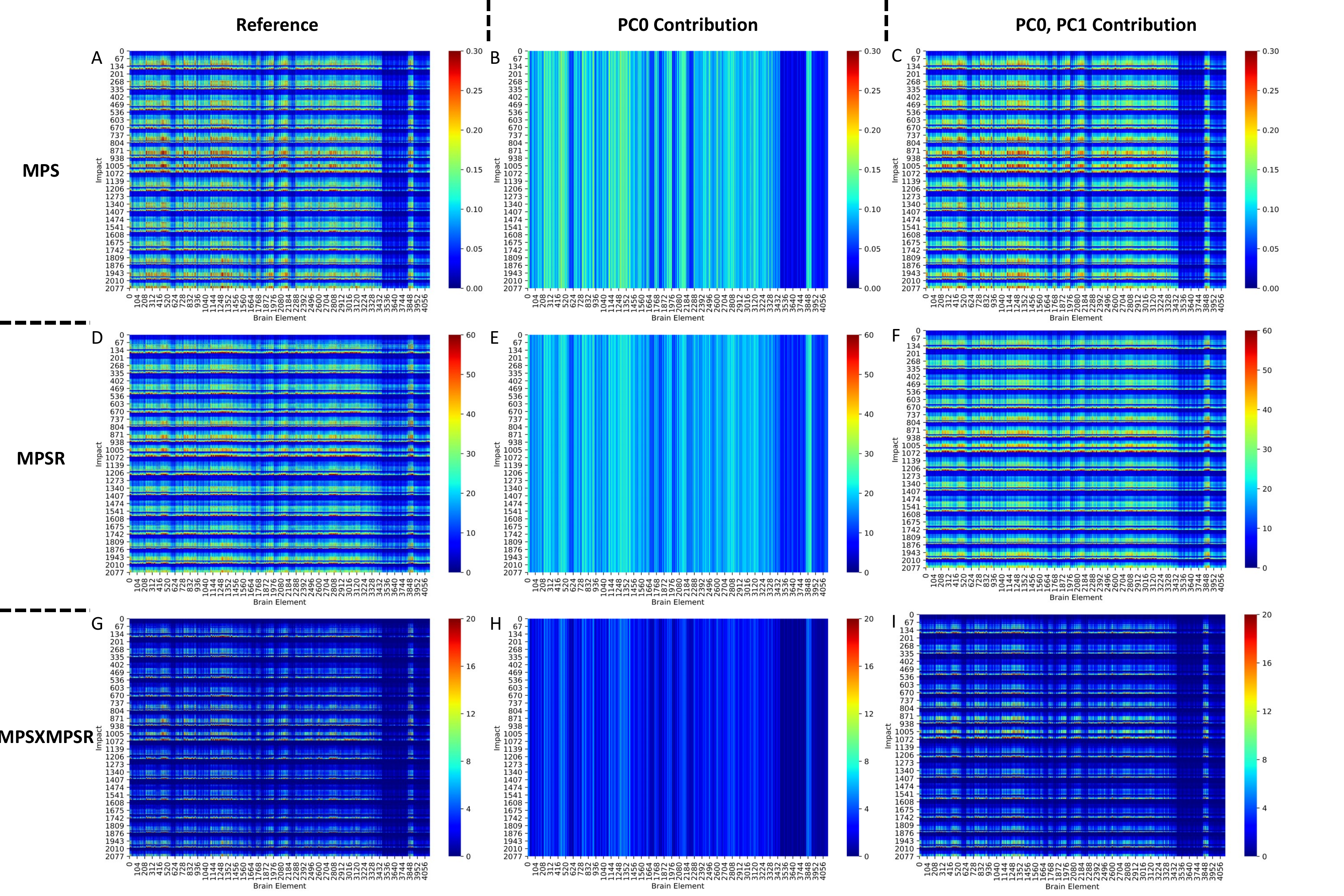}
		\caption{The heatmap visualization of the reduced-order brain dynamics representation on dataset HM. The reference values of the injury metrics, the contribution from PC0 only and the contribution from PC0 and PC1.}
		\label{example visualization}
		\vspace{-5mm}
	\end{figure*}
	
	\begin{figure*}[htbp]
		\centering
		\includegraphics[width=0.99\linewidth]{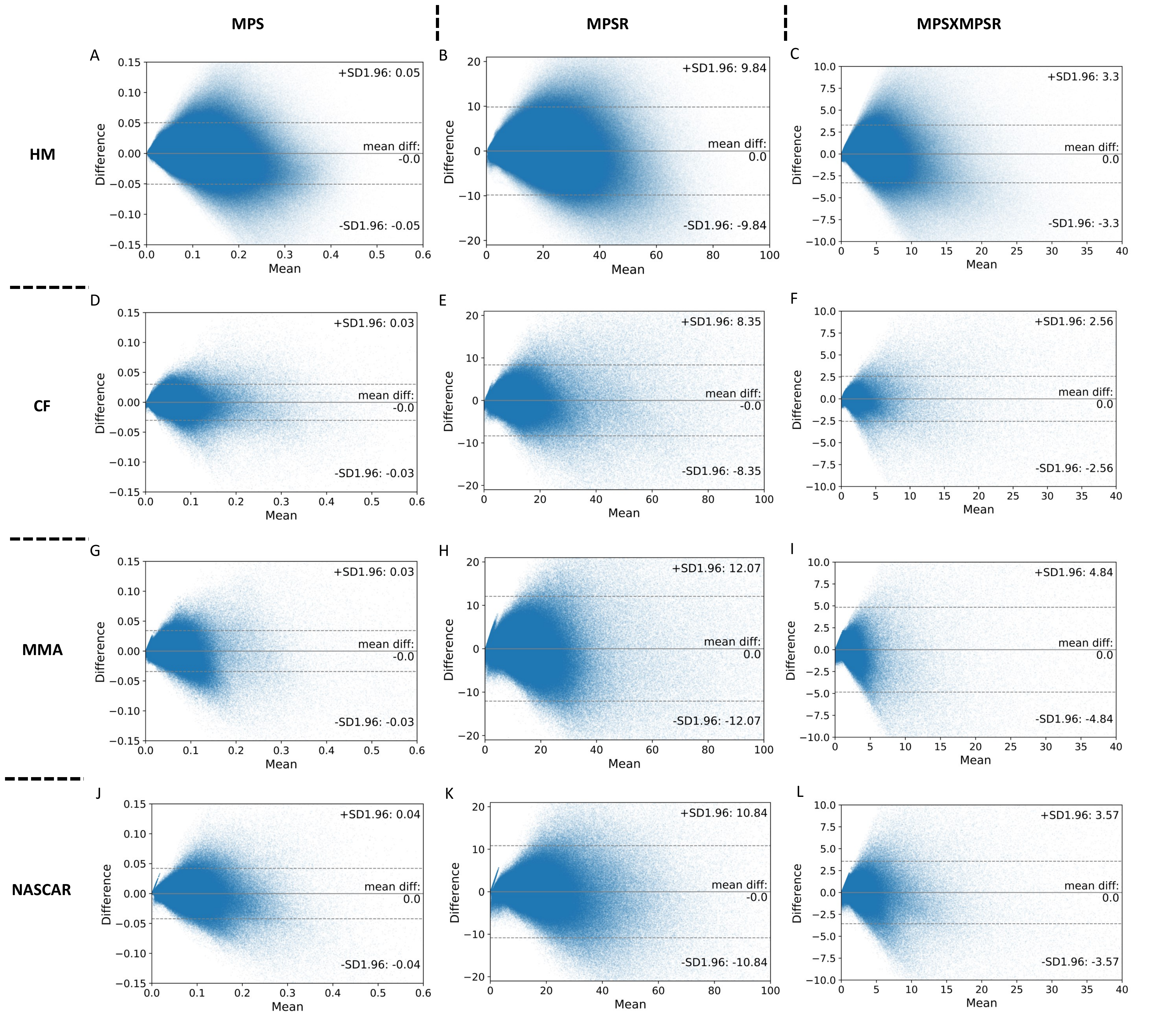}
		\caption{The Bland Altman plot of the reconstruction error with PC0 and PC1 for the three injury metrics and on four impact datasets. The x-axis denotes the mean values of the reconstructed injury metrics (with PC0 and PC1) and the reference injury metrics. The y-axis denotes the difference between the reconstructed injury metrics and the reference injury metrics (reconstructed - reference). Each dot in one subplot represents the injury metric value for one brain element and for one impact.}
		\label{reconstruction error}
		\vspace{-5mm}
	\end{figure*}
	
	\subsection{Analysis and interpretation of PC1 patterns}
\begin{figure}[htbp]
		\centering
		\includegraphics[width=0.99\linewidth]{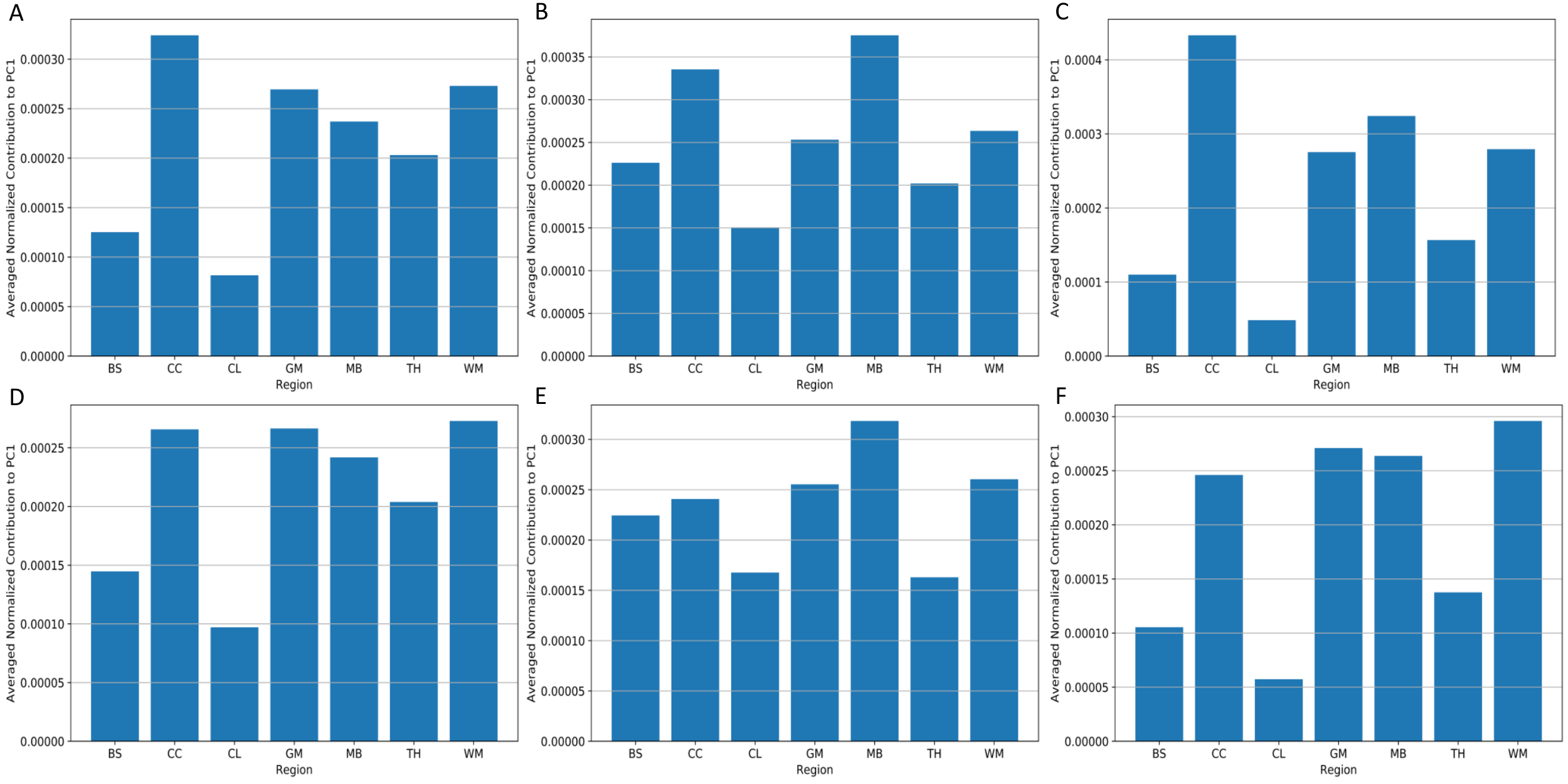}
		\caption{The averaged contribution of different brain regions to PC1 coefficients. The results come from dataset HM (A-C) and dataset CF (D-F). The contribution was firstly normalized by the sum over 4124 brain elements and then averaged over the brain elements in the same region. The different brain regions include: brainstem (BS), corpus callosum (CC), cerebellum (CL), gray matter (GM), midbrain (MB), thalamus (TH) and white matter (WM). The visualization of the PC1 coefficients for MPS (A,D), MPSR (B,E), MPS$\times$MPSR (C,F).}
		\label{PC1 Patterns}
		\vspace{-5mm}
	\end{figure}

\begin{figure}[htbp]
		\centering
		\includegraphics[width=0.99\linewidth]{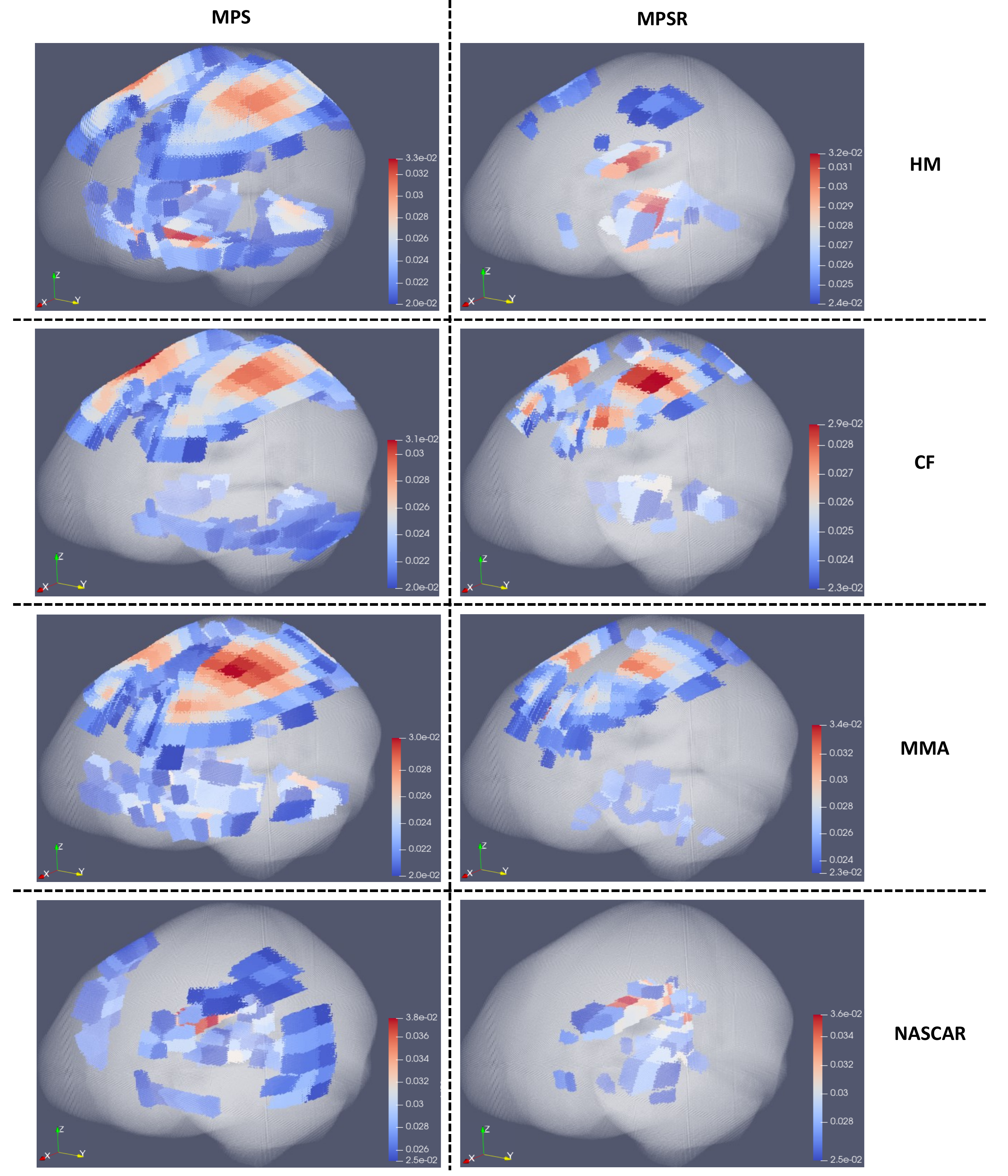}
		\caption{The 3-D visualization of PC1 coefficients for MPS and MPSR across different impact datasets. The high PC1 coefficients ($\vec{v_1}$) for MPS and MPSR are based on the entire dataset HM, CF, MMA, NASCAR, respectively. The brain elements with small PC1 coefficients are masked.}
		\label{3D}
		\vspace{-5mm}
	\end{figure}

	As PC0 and PC1 prove to explain 80\% variance in the brain dynamics, we further investigated the region-specific contribution to the PC1 coefficients (i.e., the basis vector $\vec{v_1} \in \rm I\!R^{4124}$ with the highest variance) to show the variance of injury metric in different brain regions. Because PC0 is the same across different impacts, the distribution of $\vec{v_1}$ implies the variance on specific brain elements. After normalization on $\vec{v_1}$, we extracted and averaged the values of different brain regions and visualized them in Fig. \ref{PC1 Patterns} (dataset HM and dataset CF as examples). It can be shown that the corpus callosum, gray matter, midbrain and white matter show higher contribution to $\vec{v_1}$, which indicates that the MPS/MPSR/MPS$\times$MPSR variance on these brain regions may be larger than that on brainstem, cerebellum and thalamus. To better visualize the contribution to PC1 in a 3-D matter, we plotted the high contribution regions with the low contribution regions masked (with MPS and MPSR as examples) in Fig. \ref{3D}. Based on the results, it can be shown that different impact types (i.e., football, car crashes) can lead to varying patterns in brain dynamics variation. For example, on dataset HM, CF and MMA, the cerebral cortex, corpus callosum and midbrain show higher variance in MPS and MPSR. On dataset NASCAR, the high variance occurs in midbrain and corpus callosum regions, while the cerebral cortex may not show higher variation in MPS and MPSR. 
	
	\subsection{Analysis of sensitivity of injury metrics and brain regions}
\begin{figure}[htbp]
		\centering
		\includegraphics[width=0.99\linewidth]{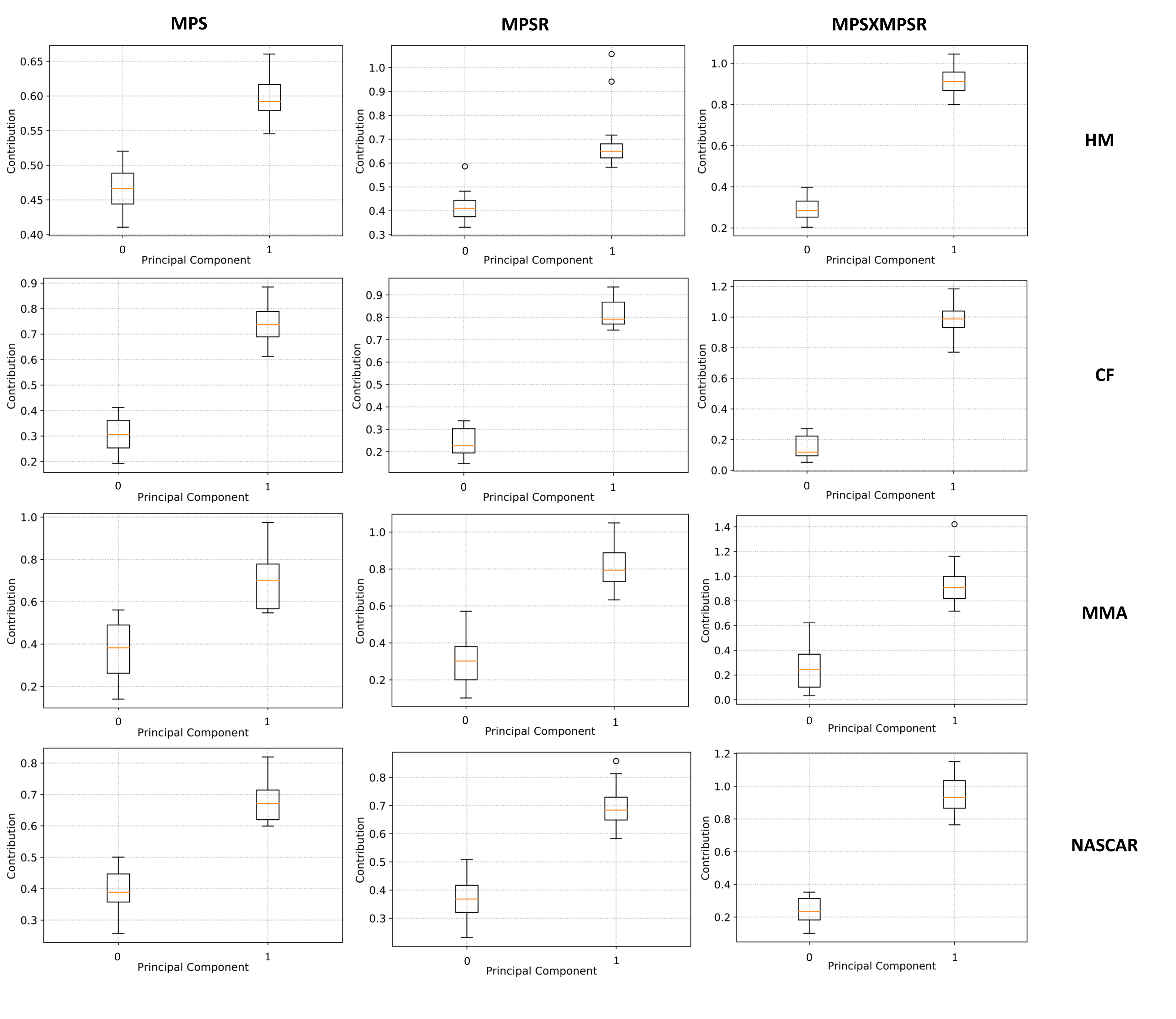}
		\caption{The contribution to the three injury metrics from PC0 and PC1 for severe impacts across four impact datasets.}
		\label{overall}
		\vspace{-5mm}
	\end{figure}

	As we found PC0 and PC1 are able to explain the majority of variance in the injury metrics on the four datasets, we then leveraged the PC0 and PC1 to summarize the brain dynamics and quantified the contribution of PC0 and PC1 in severe impacts based on the protocol introduced in Section 2C. According to the results shown in Fig. \ref{overall}, for the severe impacts across different impact types, the PC1 dominates in the contribution to the three injury metrics with statistical significance ($p<0.05$). This result indicates that generally the high MPS/MPSR/MPS$\times$MPSR values are explained by the PC1. 
	
\begin{figure*}[htbp]
		\centering
		\includegraphics[width=0.99\linewidth]{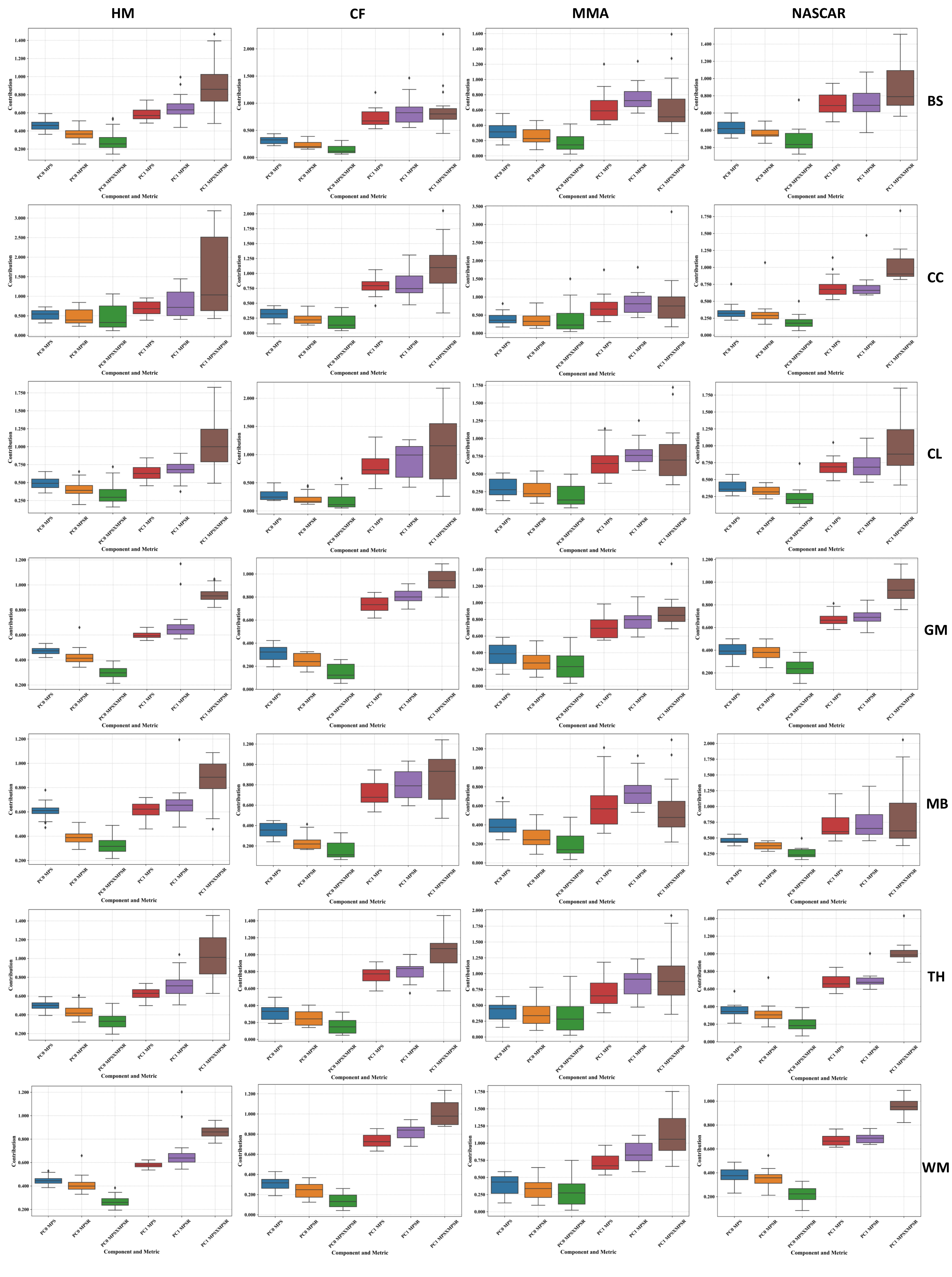}
		\caption{The contribution to the brain dynamics from PC0 and PC1 for severe impacts in the same brain region across four different impact datasets. The different brain regions include: brainstem (BS), corpus callosum (CC), cerebellum (CL), gray matter (GM), midbrain (MB), thalamus (TH) and white matter (WM).}
		\label{Dynamics Metrics}
		\vspace{-5mm}
	\end{figure*}

	Furthermore, we quantified the contribution of PC0 and PC1 in different brain regions, with results shown in Fig. \ref{Dynamics Metrics}. It can be shown that generally, the PC1 contributed more to the injury metrics for the severe impact across different brain regions and across different datasets. Additionally, for the same top 5\% impacts (shown in the boxes with the same color in each column of subplots), the corpus callosum, midbrain and cerebellum regions show higher variance in the PC1 contribution. These results indicate that for the severe impacts, these brain regions bear higher sensitivity in their brain dynamics: the brain injury metrics in these brain regions show large variations among the same top 5\% most severe impacts.
	
	Besides the sensitivity across different brain regions, it can be shown that for the same brain region and on the same dataset, the PC1 contribution to MPS$\times$MPSR shows the highest variance when compared with the PC1 contribution to MPS and that to MPSR. The PC1 contribution to MPS$\times$MPSR is also the highest. These results indicate that the MPS$\times$MPSR bears the highest sensitivity among the three injury metrics: MPS$\times$MPSR shows a large variation among the 5\% most severe impacts, and the 5\% most severe impacts deviate more from the mean. Therefore, it may be used to better differentiate the impact severity as an effective TBI predictor.
	
	\subsection{Assessment of deep learning head model with PCA}
\begin{figure}[htbp]
		\centering
		\includegraphics[width=0.99\linewidth]{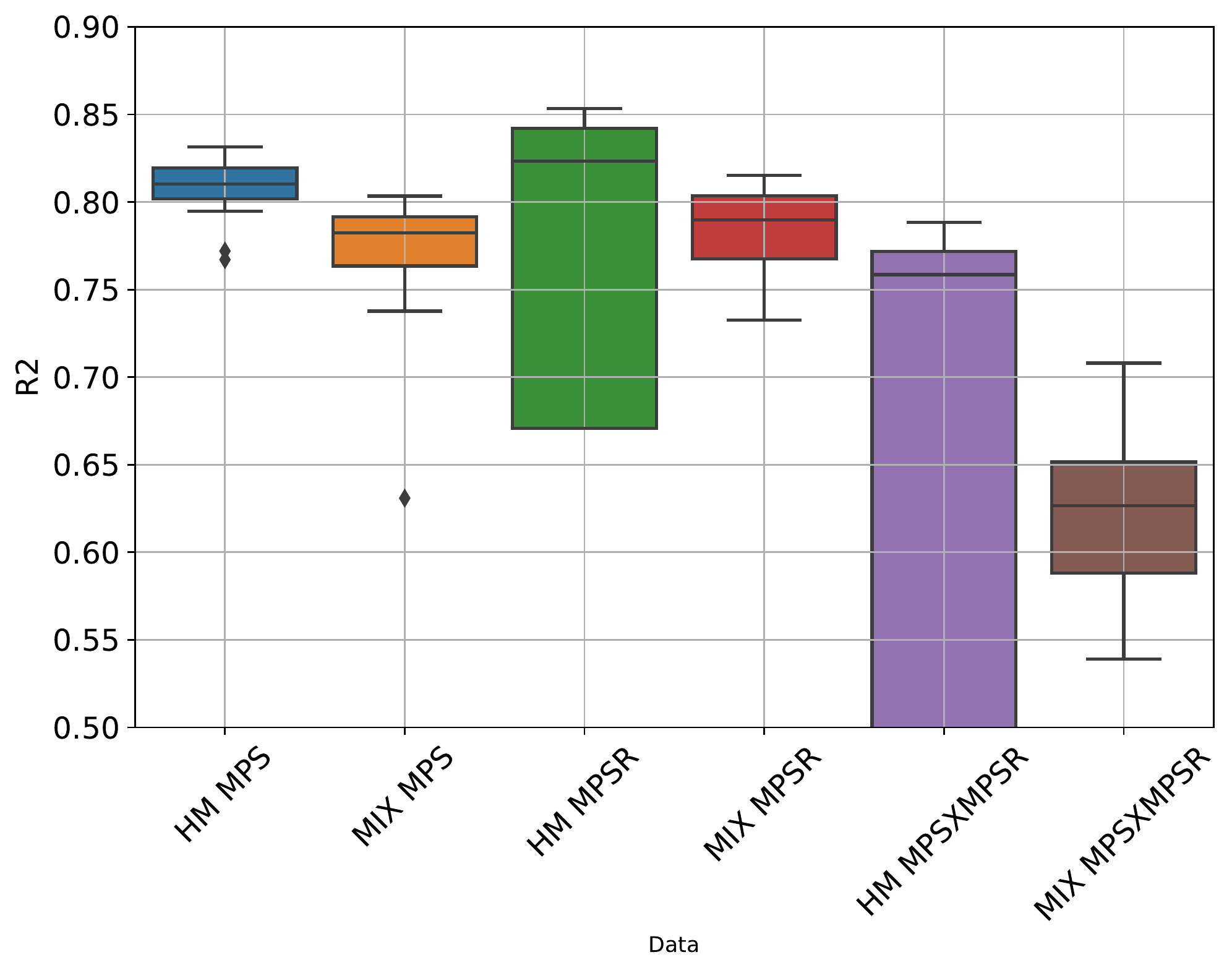}
		\caption{The accuracy of predicting the whole-brain injury metrics in terms of $R^2$. The prediction was done on dataset HM and dataset MIX, respectively. The box plots show the results over 20 parallel experiments with random dataset partitions.}
		\label{prediction}
		\vspace{-5mm}
	\end{figure}

	Based on the reduced-order brain dynamics representations with PCA, we further validated its effectiveness by developing DLHMs to predict the values on PC1 (i.e., $\vec{\alpha_1}$) from kinematics features and then reconstructed the whole-brain MPS/MPSR/MPS$\times$MPSR. The $R^2$ between the predicted values of the injury metrics and the reference values are shown in Fig. \ref{prediction}. The accuracy in terms of $R^2$ values was comparable with the DLHMs previously developed \cite{ghazi2021instantaneous,zhan2021rapid}, which indicates that predicting the values on PC1 is enough to accurately predict the whole-brain strain and strain rate. Additionally, as we visualize the distributions of the predicted values given by the PCA-based DLHMs and the reference KTH model in Supplementary Fig. \ref{prediction_distribution} (A)-(C), it is shown that the PCA-based DLHMs can generally reproduce the similar overall distributions of the MPS, MPSR, MPS$\times$MPSR when compared with the reference values. However, the prediction distributions are still slightly different from those given by the reference KTH model, which may be caused by the omission of the higher-order principal components. Furthermore, according to the error box plots in Supplementary Fig. \ref{prediction_distribution} (D)-(F), the median MAE (over 20 parallel experiments with random dataset partitions) was smaller than 0.018 for MPS prediction, smaller than 3.7$s^{-1}$ for MPSR prediction and smaller than 1.1$s^{-1}$ for MPS$\times$MPSR prediction on both dataset HM and dataset MIX. The MAE for MPS and MPSR predictions was much smaller than the presumed human concussion thresholds (~0.3 for MPS and ~25$s^{-1}$ for MPSR\cite{patton2015biomechanical,ho2007dynamic,kleiven2007predictors}) and the thresholds for accurate brain contusion volume prediction in rat model (0.3 for MPS and 2500$s^{-1}$ for MPSR reported by \cite{donat2021biomechanics}).
	
	\begin{figure}[htbp]
		\centering
		\includegraphics[width=0.99\linewidth]{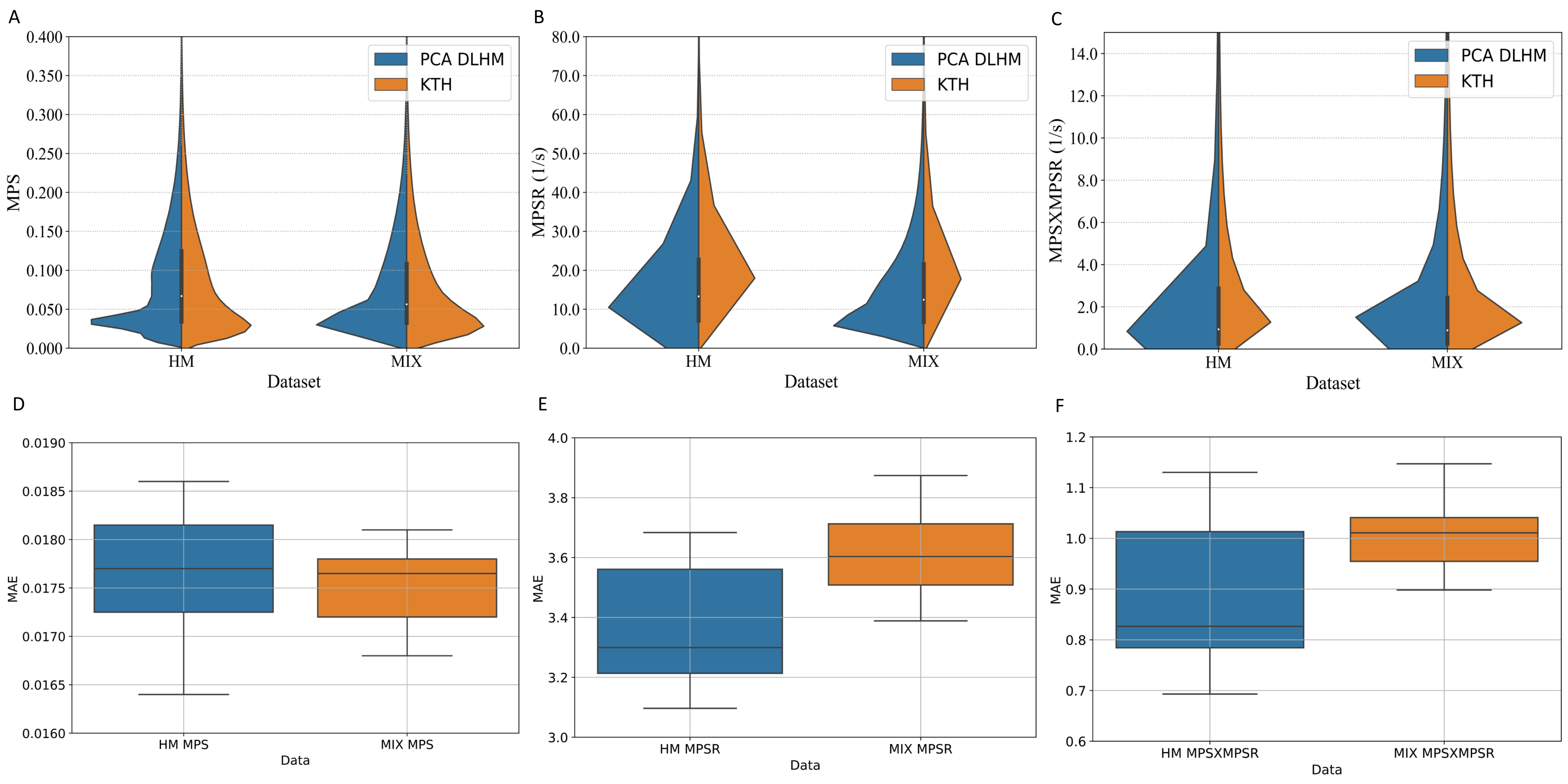}
		\caption{The overall distribution violin plots of the predictions and reference values and the mean absolute error (MAE) in the predictions. The violin plot of the overall distribution of predicted and reference MPS (A), MPSR (B) and MPS$\times$MPSR (C) on the dataset HM or the mixture of all datasets. The MAE between the predicted values and the reference values given by KTH model for MPS (D), MPSR (E) and MPS$\times$MPSR (F) on the dataset HM or the mixture of all datasets.}
		\label{prediction_distribution}
		\vspace{-5mm}
	\end{figure}

	\section{Discussion}
	In this study, we use PCA to analyze the variation of three injury metrics (MPS, MPSR and MPS$\times$MPSR) which represent the severity of brain tissue deformation during head impact, in four different datasets including simulated headform impacts (HM), college football (CF), mixed martial arts (MMA) and National Association for Stock Car Auto Racing (NASCAR). We found that the first order of principal component (PC1) is able to explain more than 80\% variation in every dataset in terms of coefficient of determination ($R^2$). The concentration of the explained variance in PC1 indicates that characteristics of the injury metrics distribution can be investigated through PC1 and the mean component (PC0). For all four datasets, we found that corpus callosum and midbrain exhibits high variance, and high variance in cerebral cortex was only found in HM, CF and MMA. In each dataset, since PC0 is the mean component and is the same for all impacts, the distribution of PC1 decides the high brain strain region that occurs frequently in the dataset. To describe how PC1 decides the injury metric in severe head impacts, we defined the contribution of the components as the predicted value by the principal component divided by the true value and validated that the contribution of PC1 dominates the injury metrics in severe impacts for every dataset. Based on the contribution of PC1, the variance of MPS, MPSR and MPS$\times$MPSR was compared at different brain regions. MPS$\times$MPSR was found to have the largest variances in severe impacts, and the MPS$\times$MPSR in the severe impacts deviates more from the mean than MPS and MPSR. Furthermore, a potential application of this study is developing models to predict the MPS, MPSR and MPS$\times$MPSR \cite{ghazi2021instantaneous,zhan2021rapid}. One of the recent efforts to improve FE head models is to incorporate small detail geometry features, which significantly increases the number of output elements and makes it difficult for modeling. Applying PCA will take advantage of the spatial relationship among elements (i.e., the co-variation among different brain elements) and reduce the dimension of prediction output. Therefore, we applied PCA to simplify our previously developed DLHM \cite{zhan2021rapid} and showed that the simplified models can accurately predict the MPS, MPSR and MPS$\times$MPSR.
	
	Because of different impact conditions (e.g., helmeted or unhelmeted, head to head or punch), the head kinematics and the relationship between head kinematics and brain deformation varies largely across different impact datasets \cite{zhan2021predictive,zhan2021relationship}. As a result, the brain regions that often experience severe deformation also vary among the types of impacts. Performing PCA on the data matrix, whose two dimensions are elements and impact, provides a tool to simplify the analysis of the spatial distribution of the injury metrics. Since PC1 is able to explain more than 80\%  of the variance of the brain (Fig. \ref{explained variance}), just PC0 and PC1 are able to represent the whole dataset (Fig. \ref{example visualization}). Since PC0 is the mean component and the same for every impact, the spatial distribution of injury metrics in severe head impact is mainly decided by PC1. Therefore, the PC1 vector is an important indicator to show the vulnerable (with high injury metric) brain regions for specific types of head impacts. For the HM, CF, MMA and NASCAR head impacts, the PC1 is visualized as a 3-D plot of the brain (\ref{3D}), and high values were found at the corpus callosum, cerebral cortex and midbrain, which indicate that these regions have a higher risk of high MPS or MPSR. This finding agrees with the previous studies comparing the mechanical loading on different brain regions \cite{hernandez2019lateral,laksari2018mechanistic,abderezaei2019nonlinear}, and these regions were found to change after 
	TBI \cite{mohamed2020evaluating,nye2021mild}. 
	
	The pattern of PC0 and PC1 in data-driven decomposition of brain dynamics suggests the discrepant loadings on different brain regions. As the injury metrics (MPS, MPSR and MPS$\times$MPSR) can be decomposed into PC0 and PC1, it is possible to further analyze the patterns of brain regions with high values of injury metrics: on the one hand, if the specific brain regions are with relatively high PC0 contribution and relatively low PC1 contributions, the regions may suffer from high strain/strain rate more constantly because the high strain/strain rate come from the mean part from the PC0 contribution. On the other hand, if the specific brain regions are with relatively low PC0 contribution and relatively high PC1 contributions, the brain regions may suffer from high strain/strain rate occasionally rather than constantly because the high strain/strain rate contribution come from the variance part from the PC1 contribution. By analyzing the patterns across different types of impacts, we are also able to observe the patterns related to different impact types. For example, MMA may constantly cause high strain on certain brain regions and occasionally cause high strain on certain brain regions. The different patterns of high strain/strain rate may also be associated with the injury patterns, and neuroimaging and histopathology data are needed to further explore this. Furthermore, since the whole-brain injury metrics can be resolved onto the PC1, PC1 scores (i.e., the elements of $\vec{\alpha_1}$) has potential to be used as an injury metric as well. The PC1 scores, which are different in different impacts, are the reduced-order summary of the whole-brain injury metrics, which may represent the overall risk better than the 95th percentile values \cite{hajiaghamemar2020embedded}.
	
	Researchers have put effort into finding the patterns of brain deformation under inertial force loading and identifying the degrees of freedom needed to accurately represent the brain dynamics under inertial force. For example, the dynamic mode decomposition was adopted to extract the deformation mode \cite{laksari2018mechanistic}, and a convolutional-neural-network-based human head model \cite{ghazi2021instantaneous} was combined with a pre-computed brain \cite{ji2015pre} response dataset to find the effective kinematics that yields similar brain deformation for the actual kinematics \cite{ghazi2021effective}. The difference between these studies and our study is that they just focused on the spatio-temporal response of the brain, while we focus on the spatial distribution of the peak values in a group of head impacts and particularly the spatial co-variation across different brain elements for a specific dataset. Therefore, the PCA results we presented here are decided by both brain physics and the characteristics of the type of head impact. 
	
	Different mechanical metrics (MPS, MPSR, MPS$\times$MPSR) were used to indicate the risk of the pathology. In this study, we compared the variance of MPS, MPSR, and MPS$\times$MPSR based on the same dataset, and found that MPS$\times$MPSR exhibits higher variance in severe head impacts, and MPS$\times$MPSR for severe head impacts deviate more from the mean. These findings suggest the potential explanation of high sensitivity of MPS$\times$MPSR and the potentially better discrimination ability of using MPS$\times$MPSR to classify injury or non-injury \cite{hajiaghamemar2020embedded,hajiaghamemar2021multi,wu2021evaluation}
	
	The different variation patterns of different types of head impacts can be shown by PC1 coefficients $\vec{v_1}$ (shown in Fig. \ref{3D}). For example, the patterns shown by the football-like impacts (without helmet) in dataset HM and the patterns shown by the college football impacts (with helmet) in dataset CF are very different. The patterns shown in the MMA impacts (without helmets) are also quite distinct from those shown in the NASCAR impacts (with helmets). Based on the PC1 coefficients, in addition to the analysis of the PC1 coefficients across the different types of head impacts and we can further leverage the PC1 coefficients to study the effects of different types of protective gears. For example, the patterns of PC1 coefficients for the impacts with or without helmets can be investigated the PC1 coefficients, because it is possible that different types of protective helmets for American football players may exert different effects on the changing patterns of the PC1 coefficients and therefore, they may redistribute the MPS/MPSR/MPS$\times$MPSR. The analysis of the PC1 coefficients enables the visualization of the protective effect of helmets in the metrics other than the simple reduction of kinematics-based metrics (e.g., peak resultant angular velocity) and strain-based metrics (e.g., MPS95: 95th percentile of MPS).
	
	The application of PCA in the development of DLHM to estimate whole-brain MPS, MPSR and MPS$\times$MPSR is also worth further discussion. DLHMs have shown their effectiveness in reducing the computational cost associated with the state-of-the-art FEM in previous studies \cite{zhan2021rapid,zhan2021rapidly,ghazi2021instantaneous}. They are developed for different FEM and act as function approximators of FEM when the mapping between the head impact kinematics and the brain dynamics is learned through large quantities of impact data. However, although the KTH model we used in this study contains 4,124 brain elements, the recently developed FEM have larger numbers of brain elements \cite{li2021anatomically,mao2013development}. Due to the high dimensionality of the output to be modeled by the DLHMs, more parameters are generally needed in the models. Therefore, large quantities of impact data are needed to train these parameters in the DLHMs. In this study, we propose to apply the PCA to firstly decompose the brain dynamics and reduce the dimension of output for the DLHMs to model, where the value on PC1 is the only target to be predicted. The PCA incorporates the spatial co-variation into the learned PC1 coefficients and the DLHMs can benefit from the spatial co-variation. After we inverse-transform the PCA process, high accuracy in the predictions of MPS, MPSR and MPS$\times$MPSR have been achieved by the DLHMs developed in this study. However, the accuracy is still not as high as that shown in our models with transfer learning technology that further take into consideration the variation across different types of head impacts. In the future, domain regularization component analysis (DRCA) \cite{zhang2017anti}, rather than the PCA, across different types of head impacts may be leveraged to further improve the prediction accuracy if there is a remarkable mismatch between the training data and test data.
	
	Although this study provides readers with a novel data-driven brain dynamics decomposition based on PCA, there are several limitations that worth notice. Firstly, in this study, we applied the KTH FE model to compute the MPS and MPSR. Although KTH FE model has been validated with experimental data of brain-skull relative motion \cite{kleiven2006evaluation}, intracranial pressure \cite{kleiven2006evaluation}, and brain strain \cite{zhou2019reanalysis}, it is relatively limited when compared with the FE models recently developed \cite{li2021anatomically,fahlstedt2021ranking}. For example, the KTH model does not consider the gyri and sulci which may significantly affect the behavior of the FE models. Additionally, it does not involve the cerebral vasculature in the modeling, which limits the modeling to be within the parenchyma. However, it has been shown by previous studies that the cerebral vasculature could significantly affect the shear stress modeled by FE models \cite{khosroshahi2021multiscale}. Therefore, for higher fidelity in brain dynamics modeling, more advanced FE models can be adopted in the future with a similar analytical pipeline introduced by this study. 
	
	In the development of DLHMs to predict the values on the first principal component, there are several limitations that need to be mentioned. Firstly, while we focus on the modeling of the first principal component (PC1), the influences of the mean components (PC0) and other components should not be ignored. It is possible that the part of variance explained by the other principal components include more detailed information closely related to the injury outcome, which warrants further validation. For the development of DLHM, in the current model, the mean components are extracted based on the mean over the training impacts, and this practice may lead to a significant decrease in prediction accuracy if the test impacts are sampled from a significantly different data distribution. Additionally, as we did not consider the higher-order principal components besides PC1, such as PC2 and PC3, the DLHM accuracy may be capped by the predictability of PC1 (i.e., the $R^2$ has an upper bound of around 0.8-0.9). The limitation can also be visualized in Fig. \ref{prediction_distribution}, where even though the overall distributions of the predictions resemble the reference distributions, the prediction distributions are still different from the reference distributions. The incorporation of more higher-order principal components may enable more accurate predictions. In the future, to enable even more accurate DLHMs based on the brain dynamics decomposition approach, higher-order principal components may be included as the targets of modeling as well, with the loss function weighing the PC1 more during the training of the DLHM.
	
	Secondly, we adopted the feature engineering approach to extract features from the temporal signals of head impact kinematics and then used them as the input of the DLHMs, considering the potential temporal mismatch of peaks and starting points of the impacts based on different measurement devices. In the future, with more data collected from various types of measurement devices, completely data-driven modeling based on convolutional neural networks \cite{ghazi2021instantaneous} and recurrent neural networks with long-short term memory (LSTM) may be leveraged to extract features from the signals in a data-driven manner, which may even lead to higher accuracy in predicting the PC1 values and enable better whole-brain MPS and MPSR predictions.
	
	\section{Conclusion}
	In this study, we applied the PCA to decompose the patterns in three brain injury metrics (MPS, MPSR, MPS$\times$MPSR) based on the spatial co-variation in four different types of head impacts. The reduced-order representation of the injury metrics enables better interpretation of the patterns in brain injury metrics across different impact types and reduces the dimensionality for deep learning head models (DLHMs) development. According to the results, we found the mean component (PC0) and the first principal component (PC1) are able to explain more than 80\% variance in most datasets. Then, we investigated the distributions among different brain regions and different injury metrics with the PC1. We showed that the corpus callosum and midbrain manifest high variance on all datasets and the MPS$\times$MPSR was the most sensitive metric to differentiate severe impacts. Finally, we leveraged the PCA to develop DLHMs to predict the three injury metrics and reached high accuracy on the predictions. 
	
	\section{Acknowledgement}\label{sec:acknowledgement}
	This work was supported by the Department of Bioengineering, Stanford University and the Office of Naval Research Young Investigator Program (N00014-16-1-2949).
	
	\bibliographystyle{IEEEtran}
	\bibliography{reference}
\end{document}